# Nanoparticle ordering in sandwiched polymer brushes

Tine Curk,[†] Francisco J. Martinez–Veracoechea,[†] Daan Frenkel,[†] and Jure Dobnikar[*,†,‡]

*University of Cambridge, The University Chemical Laboratory, Lensfield Road, CB2 1EW, Cambridge, UK, and Jožef Stefan Institute, Jamova 39, 1000 Ljubljana, Slovenia*

E-mail: jd489@cam.ac.uk

## Abstract

The organization of nano-particles inside grafted polymer layers is governed by the interplay of polymer-induced entropic interactions and the action of externally applied fields. Earlier work had shown that strong external forces can drive the formation of colloidal structures in polymer brushes. Here we show that external fields are not essential to obtain such colloidal patterns: we report Monte Carlo and Molecular dynamics simulations that demonstrate that ordered structures can be achieved by compressing a 'sandwich' of two grafted polymer layers, or by squeezing a coated nanotube, with nano-particles in between. We show that the pattern formation can be efficiently controlled by the applied pressure, while the characteristic length–scale, i.e. the typical width of the patterns, is sensitive to the length of the polymers. Based on the results of the simulations, we derive an approximate equation of state for nano-sandwiches.

---

[*]To whom correspondence should be addressed
[†]University of Cambridge, The University Chemical Laboratory, Lensfield Road, CB2 1EW, Cambridge, UK
[‡]Jožef Stefan Institute, Jamova 39, 1000 Ljubljana, Slovenia

Our ability to predict the behaviour of nanoparticles (NPs) inside polymer layers[1] is relevant for a better understanding of many biological systems,[2–5] where nanoscale objects such as proteins or viruses can interact with coated surfaces, but also for the design of inanimate nano-structured materials,[6–9] where polymer nanocomposites play an increasingly important role. In such structures, the polymers can be used to control the effective interactions between NPs and thus govern their assembly into functional structures. Conversely, NPs can alter the physical properties of the polymer layers, thus creating novel stimulus-responsive materials.[10] Examples are tunable plasmonic nano-structures,[11] ultrafast switches and organic memory devices,[12] and advanced photovoltaic devices.[13,14] Progress in the development of polymer-nano-particle materials is hampered by our lack of theoretical understanding of collective ordering of NPs in grafted polymer layers. Despite recent progress,[15–19] the quantitative prediction of the nano-structure morphology and of the kinetics of their formation remains poorly understood.[20,21] Recently, we reported Monte Carlo simulations of polymer–insoluble particles that were driven into polymer brushes by external fields.[22] The competition between entropy–favored macrophase separation and the elastic deformation of the grafted layer results in microphase separation and formation of non-uniform structures ranging from isolated clusters to percolated patterns. Such pattern formation could be controlled by varying the grafting density or the pulling force. In practice, the experimental observation of the collective organization of micron-sized colloids in brushes is complicated by the fact that preparation of several multi-micro-meter polymer brushes is challenging. This drawback would not apply to nano-sized colloids. However, in that case it is difficult to achieve a large enough external force. For instance, gravitational forces are usually too weak, and electric or magnetic field would – if they can be made strong enough at all – induce additional direct NP-NP interactions that are difficult to control. In order to observe the collective ordering of NPs in polymer brushes in experiments, it is desirable to use other techniques to insert the colloids in the polymer layers.

Here we propose a simple solution: NPs confined in a "nano-sandwich" of two polymer-grafted layers[23] will – under sufficiently strong externally applied mechanical pressure – experience qualitatively the same free energy landscape as particles pulled into a single brush by an external force.

We performed equlibrium Monte Carlo (MC) and Molecular Dynamics (MD) simulations in order to explore insertion of a single NP, as well as the collective behaviour of concentrated NPs in polymer nano-sandwiches. Depending on the polymer length, NPs density and the applied external pressure, the particles can be found in a disordered state in the middle of the sandwich, adsorbed onto the surfaces, or they form layer–spanning structures arranged in lateral patterns. We derived an approximate equation of state for the nano-sandwiches and showed that the characteristic length–scale in the systems mainly depends on the polymer length. Finally, we studied nano-structuring in curved geometries, i.e. within coated tubes and vesicles, where patterns with similar characteristic length–scale emerge upon compression.

We consider systems of polymer chains end-grafted to parallel flat surfaces separated by a distance $D$ (each polymer is end-grafted to one of the surfaces). Polymers are immersed in a neutral good solvent and are mixed with NPs (hard-sphere colloids with diameter $\sigma$ that is chosen as the unit of length in our simulations). The polymer grafting densities $\rho \approx 1/\sigma^2$ are such that without the inserted NPs the system is in–between the brush and the mushroom regime.[24,25] The relevant length scale in grafted polymer layers is the radius of gyration $R_g$ in the dilute "mushroom" regime. In the brush scaling regime it is the so-called de Gennes blob size, which is the mean distance between the grafting points. All our simulations have been performed in the intermediate regime where the two length-scales are similar and are of the order of particle diameter $\sigma$. Colloids are assumed to interact as hard spheres, i.e. they cannot overlap, while polymers are modelled as self-avoiding random walks using a coarse-grained model.[26] Each polymer chain is represented by $l_p$ soft repulsive blobs with radius of gyration $r_b = \sigma/6$ that are connected via harmonic springs $U_{ch} = 0.534\,k_BT\,(r/r_b - 0.730)^2$, with $k_BT$ the thermal energy and $r$ the center-to-center distance. The blob-blob interaction is described as a Gaussian repulsion $U_{bb} = 1.75\,k_BT\,e^{-0.80(r/r_b)^2}$, while the colloid-blob or surface-blob interaction is modelled as an exponential repulsion: $U_{bc} = 3.20\,k_BT\,e^{-4.17(r/r_b - 0.50)}$. Using this coarse-grained model, we have performed MC simulations and evaluated the single–particle insertion free energy profiles following the Wang-Landau approach[27] described in ref.[22] Fig. 1 shows the free energy profile for a single NP, confined inside a polymer sandwich. When the two coated surfaces are well separated, the free energy has a global minimum at the center and metastable minima at the surfaces separated from the global minimum by a barrier. Interestingly, the shape and heights of the barrier can be tuned by compressing the sandwich: upon sufficient compression the stability is inverted, i.e. the surface minima become global and the central one metastable.

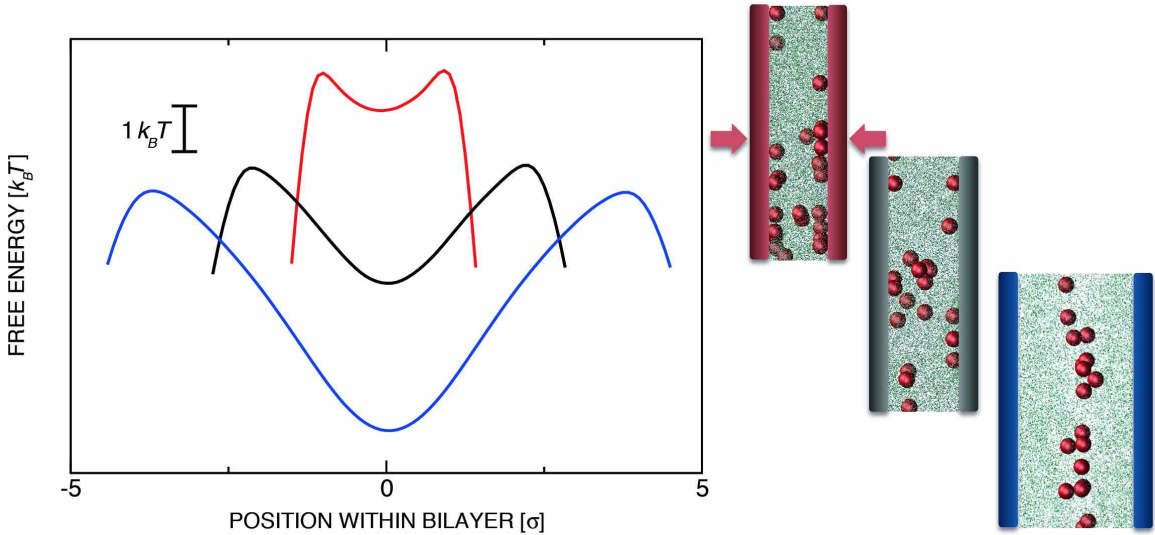

Figure 1: The free-energy landscape for a single NP confined between two polymer–grafted surfaces with disordered grafting. At large separations the global free energy minimum is in the middle between the plates and the minimum at the surfaces is metastable. Upon compressing, the minimum at the surface becomes stable, which means that the NPs will stick to the walls. At the right-hand side, the NP-ordering corresponding to the three typical free–energy landscapes is shown, as obtained by MC simulations in dilute sandwiched system. NPs are represented with red spheres and the polymers are shown as a series of green dots.

These results are applicable to the dilute regime with a few inserted particles that do not significantly interact. As observed in MC simulations (snapshots on Fig. 1), NPs concentrate in the interfacial region in the middle of the sandwich when the surface separation is large. However, at sufficiently small separation - when larger pressures are applied - they adsorb to the two surfaces. Thus, by pressing nano-sandwiches together, NPs that were originally adsorbed on the polymer brushes, become trapped at the surfaces. If we then separate the two surfaces by releasing the pressure, the free–energy landscape changes (see Fig.1) and the surface-adsorbed NPs are trapped in metastable minima. Their subsequent release into the bulk is a slow process controlled by the height of the free–energy barrier. Such processes are size–selective and biased towards adsorption/release of smaller NPs, which is important in many biological systems where intrusion of large particles has to be inhibited. The barrier height, an thereby the colloidal release rate, can be controlled by varying the grafting density $\rho$.[22]

In the concentrated regime the NPs interact via polymer–mediated many–body interactions that govern their self–assembly. We have performed MC simulations of dense colloidal sandwiches in the $NVT$ ensemble. In order to achieve NP organization into collective nanostructures, the magnitude of the polymer–induced NP–NP interactions needs to be larger than $k_BT$. This magnitude depends predominantly on the effective 3D density of monomers in the space around NPs that is accessible to the polymers. A sufficiently high monomer density can be achieved by having long polymers, high grafting density, or a small sandwich thickness. In principle, similar morphologies are stable if we consider shorter and denser or longer and less densely grafted polymers. However, increasing the grafting density $\rho$ has a significant effect on the kinetics of structure formation. In order to avoid kinetic arrest, we have limited its values to $0.5 \leq \rho\sigma^2 \leq 1$ and explored the behaviour of the system by varying the other parameters: $l_p$, $D$ and number of particles $N_p$. Fig. 2 displays typical simulation snapshots at representative parameter values. The number of blobs per chain is set to $l_p = 60$. At large plate separation ($D = 8.3\,\sigma$) the NPs are mostly located in the interfacial region between the plates. Upon compressing the sandwich the NPs protrude from one plate to another and form layer-spanning structures ($D = 5.8\,\sigma$). These are either isolated tower-like columns in a relatively dilute case or laterally percolating structures in a dense NPs regime.

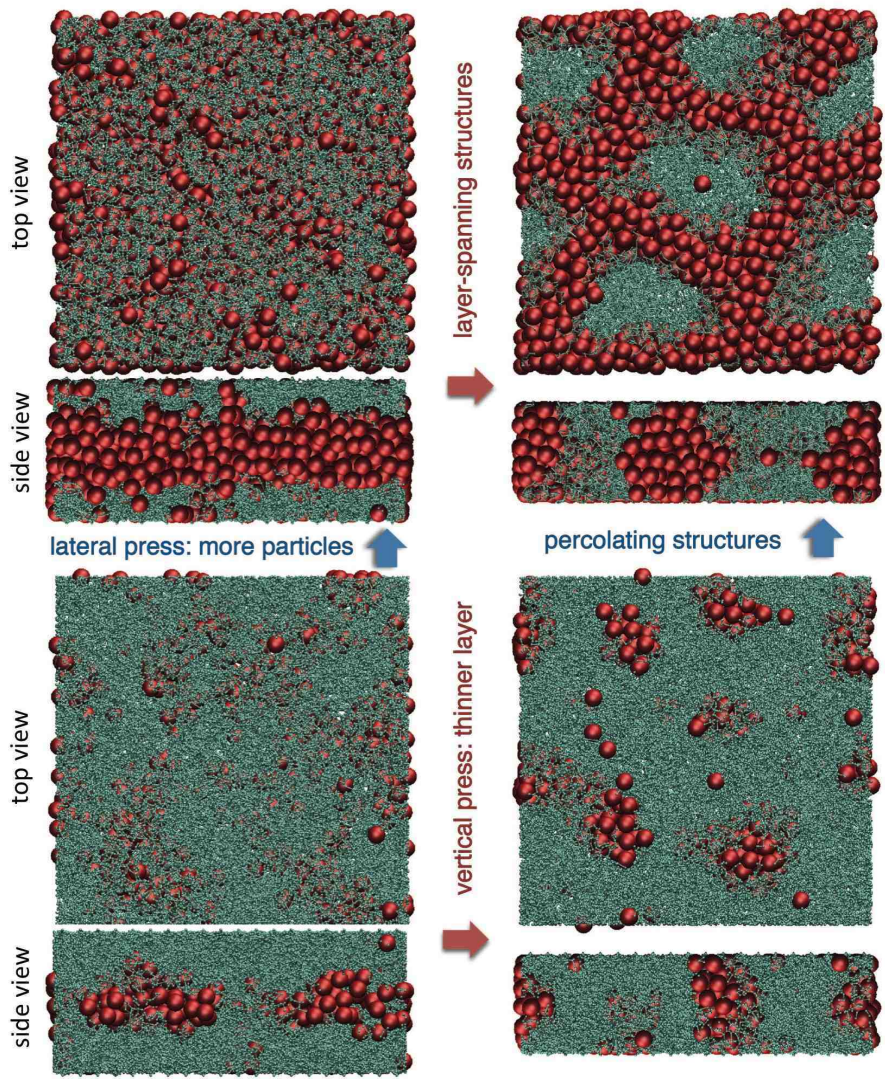

Figure 2: Top-view and side-view snapshots from canonical Monte Carlo simulations for 400 (lower row) and 1400 (upper row) NPs in sandwiched brushes with plate-to-plate separation $D = 5.8\sigma$ (right column) and $D = 8.3\sigma$ (left column). The simulation box size is $20\sigma$ x $20\sigma$ in all cases, with periodic boundary conditions in lateral directions. We assume that the number of NPs in the system could be controlled by varying lateral pressure via barostat. In order to ensure that the polymer-induced interactions are strong enough to cause collective ordering the grafting density is larger in the dilute case with ($\rho = 1.0$ for 400 and $\rho = 0.5$ for 1400 particles). At large plate separation the NPs predominantly occupy the interstitial region between the plates. Upon compressing the plates they form layer-spanning structures connecting both plates. An efficient control of the NPs ordering could thus be experimentally achieved by controlling $N_p$, e.g. by a barostat, and the pressure $p$. NPs are represented with red spheres and the polymers are shown as a series of connected green dots.

The isolated clusters are stabilised and do not coalesce because the local density of the polymers is increased around the clusters. The characteristic length–scale of all the observed patterns can be understood in terms of the competition between polymer-particle phase separation and the elastic–like deformation of the grafted polymers. The mathematical theory of the structure formation developed for single polymer layers [22] can be successfully applied to the system of sandwiched brushes studied here.

In order to characterize the observed patterns we have performed additional constant–pressure ($NpT$) MD simulations using a standard velocity Verlet integration scheme coupled to a Lowe thermostat.[28] The length of the time step was chosen so that in a single simulation step a polymer blob moves by roughly a fraction $1/20$ of its radius of gyration. A constant–pressure simulation was implemented in a mechanistic way: one of the plates was kept fixed at $z = 0$ and a constant pressure was applied to the other plate. The model for polymers used in the MD simulations was identical to the one from the MC simulations, and the hard–core repulsion between NPs and between NPs and walls was modelled by the standard WCA potential,[29] see Online SI for details.

The system with lateral dimensions $40\sigma$ x $40\sigma$ was initially prepared as a well–separated sandwich brush. We then slowly increased the applied pressure, causing $D$, the gap between the hard walls, to decrease. The pressure – measured in units of $p^* \equiv k_B T/\sigma^3$ – was linearly increased from 4 to 26 in $10^7$ time steps, which is slow enough so that the polymer matrix was in a quasi equilibrium during the simulation. This claim is supported by the fact that we observed no hysteresis in the equation of state $D(p)$ when we subsequently decreased the pressure back to the initial value (see Online SI). The time evolution of the system upon increasing the pressure depended on the particle load in the polymer layer, $N_p$. In dilute systems ($N_p \lesssim 200$) we recover the transition from the disordered gas–like state in the middle of the sandwich to adsorption of the NPs at the surfaces in compressed sandwiches. If the density of NPs is larger, they initially form a continuous liquid–like state, which becomes unstable upon compression and breaks up into smaller clusters – still in the middle of the sandwich. Upon further increasing the pressure, the clusters elongate in the $z$ direction and form layer-spanning structures, either isolated towers at small $N_p$, or percolated networks in dense regime. Typical lateral patterns are displayed in Fig. 3**A)** for two different values of the polymer length $l_p$. The patterns can be characterized by a length–scale $L$, which describes either the diameter of the towers, or the width of the stripes. We can determine $L$ by inspection of simulation snapshots or, more quantitatively, by determining the first peak in the static structure factor (see Online SI). We note that $L$ clearly correlates with $l_p$, see Fig. 3**B)**, while it hardly varies with the particle load $N_p$ (inset of Fig. 3**B)**). We must remark that in principle $L$ depends also on the grafting scenario: at the same average grafting density it is slightly larger on disordered surfaces where local density fluctuations at the surface are appreciable, and smaller for uniformly grafted ordered surfaces. In any experimental realization the grafting is likely to be between the completely disordered and completely ordered. Here we only show the results for ordered grafting, see Online SI for comparison to the disordered grafting and a discussion.

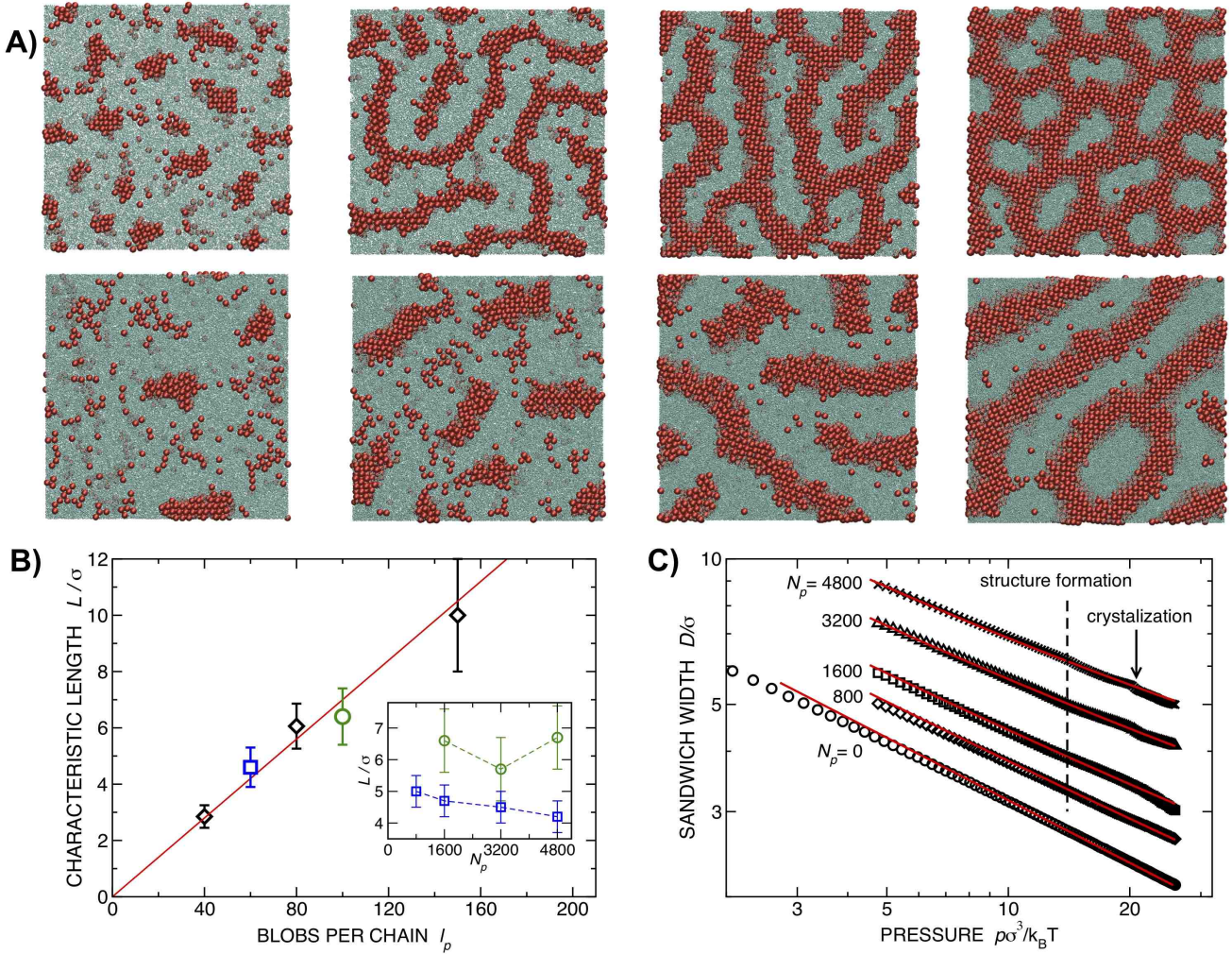

Figure 3: **A)** Lateral patterns in systems with different polymer length (top row: $l_p = 60$, bottom row: $l_p = 100$) and number of particles (columns from left to right: $N_p = 800, 1600, 3200$ and $4800$). The simulation box size is $40\sigma \times 40\sigma$ with periodic boundary conditions in lateral directions. The snapshots were obtained in MD simulations at a fixed pressure $p = 25 k_B T/\sigma^3$. **B)** Dependence of the characteristic length–scale $L$ on the polymer length $l_p$ at fixed number of inserted nanoparticles $N_p$ (main plot) and on $N_p$ at fixed $l_p$ (inset). The length-scale $L$ was obtained from the first peak of the static structure factor (see Online SI). In the main plot, the value of $N_p = 3200$ was used and in the inset the two curves correspond to $l_p = 60$ (blue symbols, lower curve) and $l_p = 100$ (green symbols, upper curve). The lines are guide to the eye. **C)** The equation of state for nano-sandwiches obtained by MD simulations (symbols) and from Eq.(1) (red solid lines) with $\omega = 0.43, K = 8.55$ and hard–sphere equation of state for the nanoparticles.

In Fig. 3**C)** the 'equation of state', i.e. the computed dependence of the sandwich thickness $D$ on the external pressure $p$ is plotted for systems with different number of particles $N_p$. The measured equation of state can be understood in simple theoretical terms. For an empty ($N_p = 0$) polymer sandwich, the scaling arguments[24] predict $D_0 \propto p^{-\omega}$ with the exponent $\omega$ that depends on the self-avoiding walk critical scaling exponent $\nu$ as $\omega = \frac{3\nu - 1}{3\nu}$. The value of the critical exponent in three dimensions, $\nu \approx 0.588$, implies $\omega \approx 0.433$ that agrees well with the value deducted from the simulations, $\omega \approx 0.43 \pm 0.01$. The comparison between the scaling relation $D_0 \propto p^{-0.43}$ and the equation of state measured in simulations is shown in Fig. 3**C)** (lower-most curve).The agreement is very good except for low pressures where the system is not yet in the brush scaling regime and the agreement is not expected. The volume accessible to the polymers in an empty sandwich is equal to the entire volume of the system, $V_0 = SD_0 = SKp^{-\omega}$ where $S$ denotes the surface area of the hard plates. $K$ is a proportionality constant that can be estimated from the lower red curve in Fig. 3**C)**: for $l_p = 60$ and $\rho = 0.5$ the value is $K \approx 8.5$.

In order to arrive at a theoretical estimate of the effect of the nano–particles on the equation of state of the NP-polymer system, we make the following assumptions: **i)** in the regime where nano–particle structure form, the polymers and colloids are fully phase separated and **ii)** in computing the volumes of the nano–particle and polymer phases, we ignore surface effects. These assumptions imply that we describe that cluster of nano-particles as dense fluid (solid) of (almost) hard spheres, and the polymer brush is treated as a pure polymer brush occupying the volume not occupied by the nano–particles. Hence, the total volume of the system is given simply by a sum of the volumes $V(p) = V_0(p) + V^{NP}(p)$, where we denote the volume occupied by the nano–particles by $V^{NP}(p)$ and recall that the volume of the polymers at pressure $p$ is denoted by $V_0(p)$. We assume that the nano–particle clusters obey the equation of state of the WCA model and hence $V^{NP}(p) = N_p/\rho_p(p)$, where $N_p$ is the number of nano–particles and $\rho_p(p)$ is the nano–particle density. Dividing by the surface area $S$ allows us to write the following approximation for the thickness $D(p) = V(p)/S$ of the nano-sandwich

$$D(p) = D_0(p) + \frac{V^{NP}(p)}{S} = Kp^{-\omega} + \frac{N_p/S}{\rho_p(p)} . \qquad (1)$$

We note that $K$ is a proportionality constant and $\omega$ a scaling exponent as discussed above. In Fig. 3**C)** we plot the predictions resulting from Eq.(1) together with the simulation points for the planar sandwiched geometry. To obtain the theoretical curve, we have used the mapping of Heyes and Okumura[30] of the WCA equation of state onto the hard-sphere equation of state taken from.[31] We observed an excellent agreement between the simple theory and the simulation results. In principle, the equation of state (Eq. 1) would also be valid for

non–planar cases or in curved geometries – as long as the scaling exponent $\omega$, which is geometry–specific, is correctly evaluated.

We observe no discontinuity in $D(p)$. At the transition point where colloidal ordering sets in, equilibration is slow, but we find no evidence for a first-order jump in $D(p)$: slow compression and expansion revealed very little, if any, hysteresis (see Online SI). At higher values of the pressure, there is a crystallisation transition within the colloidal structures: this transition appears to be first-order and is associated with hysteresis.

From Fig. 3C) we can see that – in the reduced units – the typical pressure needed to induce the structure formation is $p/p^* \approx 12$. The value of the pressure unit $p^* = k_B T/\sigma^3$ depends on the NPs size: At room temperature, $T = 293$K, the value ranges from a few Pa for 100 nm NPs to MPa for NPs with diameters of 1 nm, see Table 1. Therefore, the physical pressure one needs to apply to induce the structure formation should be easily accessible in the experiments.

Table 1: Conversion between the reduced and physical units. The reduced pressure is measured in units of $p^* = k_B T/\sigma^3$, where $k_B$ is the Boltzmann constant and $T = 293$K is the temperature.

| NP size $\sigma$ | 1 nm | 10 nm | 100 nm | 1 $\mu$m |
|---|---|---|---|---|
| Unit of pressure $p^*$ | 4 MPa | 4 kPa | 4 Pa | 4 mPa |

We have examined the effect of additional nanoparticle–polymer interactions on the observed properties of the confined system (for details, see Online SI). In the case of attractive NP–polymer interactions, the homogeneous fluid–like state of the dispersed nano–particles is stabilized and we do not observe the formation of inhomogeneous structures of nano–particles. Such behaviour is to be expected as the attractive colloid-polymer interactions favour mixing. In case of repulsive NP–polymer interactions (beyond excluded volume), the ordering transition becomes more step-like (see Online SI). The preliminary results suggest that at sufficiently strong repulsions the micro–phase separation may become first–order. We have not explored this regime.

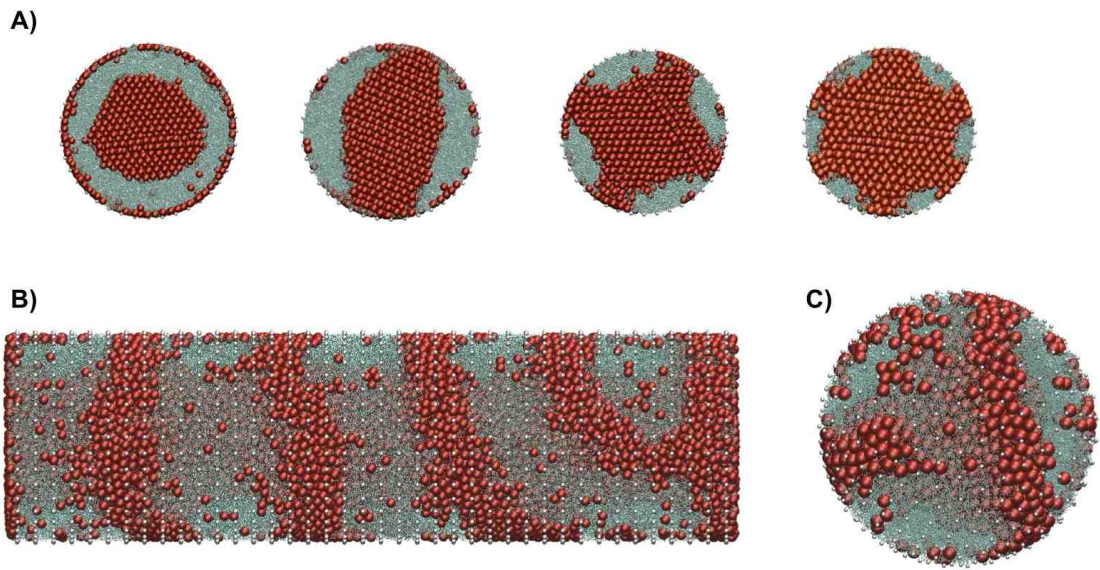

Figure 4: NP ordering in confined grafted systems with different geometries: **A)** disks, **B)** cylinders, and **C)** spheres. NPs are represented with red balls, polymers are coloured green and polymer anchoring points are represented with white spheres. The external pressure is $17 k_B T/\sigma^3$ throughout the panel. Disk confinement results in polar patterns that resemble "stars" where the number of "arms" can be controlled by the polymer length: in the snapshots **A)** the only parameter that changes is the polymer length, from left to right $l_p = 400$, 300, 200, 100. **B), C):** In a cylindrical and spherical confinement NPs form patterns with characteristic length scales similar to that in the planar geometry. Details of the simulations with complete parameter values and movies can be found in the Online Supplementary Information.

We expect that the planar slit geometry described above will be most relevant for experiment. However, we note that similar micro phase separations can be observed for nanoparticle–polymer systems in cylindrical and spherical confinement. To explore this behaviour, we performed MD simulations of NP-polymer systems in two–dimensional circular confinement (Fig. 4**A)** and observed nanostructures governed by the characteristic length scale and by the strong confinement effects. In three–dimensional nanocylindrical tubes coated on the inside (Fig. 4**B)**, we observed regular patterns reminiscent of the nanostructures formed in confined diblock copolymer melts.[32] Finally,we also considered spherical vesicles coated on the inside. In order to ensure uniform grafting density in spherical geometry, we chose the positions of the grafting points to be the solutions of the Thomson problem.[33] Typical resulting patterns are shown on Fig. 4**C)**. We have not investigated the non-planar systems comprehensively, however, we did observe that the characteristic length scale of the patterns is similar (i.e. within a factor of 2) to that of the equivalent planar system.

In conclusion, we have studied the spontaneous organization of nanoparticles in a confined, polymer–coated systems under mechanical pressure. We observe that the colloids assemble in patterns that are similar to those found in sedimenting micro–colloids. The phenomena that we predict on the basis of our simulations could be tested in experiments on nano-colloids trapped between polymer brushes. Our results suggest that it should be possible to design thin nano-structured polymer-nanoparticles mixtures with a switchable mesoscopic structure. Conducting NPs mixed with a semiconducting polymer layer have found an application in the design of photovoltaic cells.[13] In that case, the NP–phase needs to interpenetrate the polymer layer and span the space between both electrodes. Laterally, isotropic structures with a characteristic length–scale of about 20 nm are desirable[34] in order to minimize the recombination of the excited electron–hole pairs before reaching the electrodes. Our results suggest a route to control the morphology and scale of nano–composite structures for photovoltaic materials. The attractive feature of the procedure that we propose is that nano-particle structuring is enforced simply by pressing two coated surfaces together. Once the particle structures are formed within the polymer layer, the latter can be cross linked to make the structure permanent. It would be interesting to explore to what extent the formation of pressure-induced nano-colloidal structures plays a role in biological systems. A possible example might be the load–bearing cartilage in joints.

## Acknowledgement

We gratefully acknowledge enlightening discussions with Oren Scherman, Richard Friend, Tobias Kraus and Nicholas Tito. This work was supported by the 7th Framework Program of European Union through grants ARG-ERC-COLSTRUCTION 227758 and ITN-COMPLOIDS 234810, by the Herchel Smith Fund, and by the Slovenian Research Agency through Grant P1-0055.

The authors declare no competing financial interest.

## Supporting Information Available

In the supporting information we present more details about the model and simulation techniques. We also show additional results for a broader parameter range and 2D structure factors, we also include movies of the simulated system evolution. This material is available free of charge via the Internet at `http://pubs.acs.org/`.

# Graphical TOC Entry

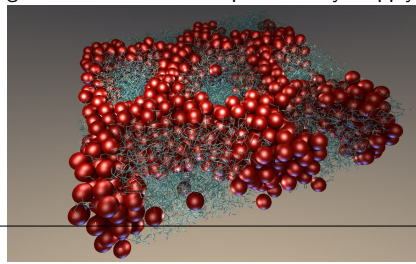

Self-organization of nanoparticles sandwiched between two polymer-grafted surfaces compressed by applying mechanical pressure.